# A Design and Analytic Strategy for Monitoring Disease Positivity and Case Characteristics in Accessible Closed Populations

Robert H. Lyles, Department of Biostatistics and Bioinformatics, The Rollins School of Public Health of Emory University, 1518 Clifton Rd. N.E., Atlanta, GA 30322, USA (phone: 404-727-1310; fax: 404-727-1370; email: rlyles@sph.emory.edu)

Yuzi Zhang, Department of Biostatistics and Bioinformatics, The Rollins School of Public Health of Emory University, Atlanta, GA, USA

Lin Ge, Department of Biostatistics and Bioinformatics, The Rollins School of Public Health of Emory University, Atlanta, GA, USA

Lance A. Waller, Department of Biostatistics and Bioinformatics, The Rollins School of Public Health of Emory University, Atlanta, GA, USA

# Abstract

We propose a monitoring strategy for efficient and robust estimation of disease prevalence and case numbers within closed and enumerated populations such as schools, workplaces, or retirement communities. The proposed design relies largely on voluntary testing, notoriously biased (e.g., in the case of COVID-19) due to non-representative sampling. The approach yields unbiased and comparatively precise estimates with no assumptions about factors underlying selection of individuals for voluntary testing, building on the strength of what can be a small random sampling component. This component unlocks a previously proposed "anchor stream" estimator, a well-calibrated alternative to classical capture-recapture (CRC) estimators based on two data streams. We show here that this estimator is equivalent to a direct standardization based on "capture", i.e., selection (or not) by the voluntary testing program, made possible by means of a key parameter identified by design. This equivalency simultaneously allows for novel two-stream CRC-like estimation of general means (e.g., of continuous variables such as antibody levels or biomarkers). For inference, we propose adaptations of a Bayesian credible interval when estimating case counts and bootstrapping when estimating means of continuous variables. We use simulations to demonstrate significant precision benefits relative to random sampling alone.

# Introduction

The importance of monitoring communities for disease, particularly in the case of vulnerable and workforce populations, is acutely highlighted by the COVID-19 pandemic. This is true from the standpoint of identifying individual cases, but also for assessing prevalence to gauge community risk. However, voluntary or partially mandated testing programs typically produce positivity data from non-representative samples of the target population, oversampling those with symptoms, suspected exposure, vulnerability concerns, etc. (1). Biases resulting from such convenience samples underscore the value of representative sampling (2), which has influenced large-scale efforts to monitor COVID-19 (3-5). Such surveys are crucial, though it is understood that the greater the scale, the more complex the challenges to implementation and validity due to issues such as non-response (6).

In smaller-scale closed communities such as retirement homes, universities, or workplaces, voluntary testing programs are potentially crucial. Using COVID-19 as an example, these programs make testing available on designated days for those seeking it, perhaps with triggers for mandatory testing, e.g., for the unvaccinated or persons exhibiting symptoms. The resulting non-representative sample may be of little use for accurately monitoring prevalence, but critical for identifying individual cases and lowering community risk. Given the smaller scale and accessibility of a complete list of community members, however, the addition of a random sampling-based component to surveillance is more likely feasible than in large open population settings. Using a recently outlined "anchor stream" design (7), our goal is to encourage occasional planned sampling to augment voluntary disease testing in such settings. While this approach identifies additional cases, its main advantage is to unlock efficient estimators of the current prevalence and case count by making use of the voluntary testing program together with the random sample implemented by design. Importantly, the former may yield the lion's share of observed cases, which are accounted for analytically regardless of the nature of their non-representativeness.

Along with the need to estimate prevalence and case counts, there is often a compelling rationale for augmenting disease monitoring by simultaneously measuring biomarkers. For example, researchers have identified associations between disease severity and elevations of biomarkers associated with influenza (8-9) and COVID-19 (10-11). With these markers typically measured on a continuous scale, such augmented monitoring would benefit from design and analytic approaches to combat the same issues with non-representative sampling that plague case count estimation. We address this need here, providing a design-facilitated unlocking of capture-recapture (CRC)-related methods to estimate general means that may offer promising benefits for disease monitoring in closed communities.

*Review of the Anchor Stream Design*

The design we promote is known to theoretically justify the classical Lincoln-Petersen (12-13) and Chapman (14) CRC estimators of a population total (N) based on two surveillance streams. Specifically, if one stream is implemented as a random sample of the target population and identifies individuals independently of capture status based on the other, the classical estimators are valid (15-17). A well-known challenge in CRC settings is that it can be very difficult know whether such Lincoln-Petersen (LP) independence conditions are approximately correct. Chao et al. (18) cite the example of post-enumeration surveys in conjunction with the census, while Lyles et al. (7) suggest a similar strategy for estimating case counts in registry-based populations or in smaller-scale closed communities.

To fix ideas, consider a retirement community featuring apartment-style living with staff devoted to health and wellness, where residents are encouraged to undergo voluntary disease testing on a periodic basis (e.g., Monday mornings). While this approach may be effective at detecting cases among those who consider themselves at high risk, such a voluntary monitoring effort ("Stream 1") undoubtedly produces a non-representative sample that cannot be relied upon to estimate the current community-wide prevalence or case count. The study design of interest is to augment Stream 1 by randomly selecting and testing a relatively small sample of residents from the population. This random sample ("Stream 2") should be selected at the close of voluntary testing on the day in question (i.e., a post-enumeration), to avoid affecting individuals' likelihood of self-selection into Stream 1. The sample could be drawn from the full list of community members (with those selected who were already tested voluntarily not requiring another test), or from the list of members after excluding those tested in Stream 1. We refer to Stream 2 as an "anchor stream", formally requiring the following conditions (7): a) Individuals are notified of random selection for testing in Stream 2 after the daily voluntary test window is closed, thus without influence of or by their inclusion in Stream 1; b) Positive and negative test results are recorded and linked to individuals in Streams 1 and 2; c) The number of individuals who are inaccessible or unwilling to participate in random testing if sampled is negligible; d) The testing method identifies cases and non-cases without error (this could be relaxed; see Discussion).

**Methods**

**Table 1** allocates the $N_{tot} = \sum_{j=1}^{7} n_j$ members of a closed target population into one of 7 cells based on execution of the anchor stream study design. As noted previously (7), the design immediately validates two simple estimators of the prevalent case count. The first is the standard estimator based on the Stream 2 random sample, i.e.,

$$\hat{N}_{RS} = N_{tot}\hat{p}_{RS} \quad,$$

where $\quad \hat{p}_{RS} = n_{RS}^+ / n_{RS} = (n_2 + n_6)/(n_1 + n_2 + n_5 + n_6), \ \text{Vâr}(\hat{N}_{RS}) = N_{tot}^2 \text{Vâr}(\hat{p}_{RS}) \ ,$

and
$$\text{Vâr}(\hat{p}_{RS}) = \left(\frac{n_{RS}(N_{tot}-n_{RS})}{N_{tot}(n_{RS}-1)}\right)\left(\frac{\hat{p}_{RS}(1-\hat{p}_{RS})}{n_{RS}}\right) \quad (1)$$

In expression (1), the premultiplying term is a finite population correction (FPC) based on an unbiased variance estimator given by Cochran (19). In the event that this premultiplying term exceeds 1, we set it equal to 1.

The second simple alternative is a traditional bias-corrected CRC estimator based on two data streams (14):

$$\hat{N}_{Chap} = \frac{(n_{1\bullet}+1)\times(n_{\bullet 1}+1)}{n_{11}+1} - 1 \quad, \quad \text{Vâr}(\hat{N}_{Chap}) = \frac{(n_{1\bullet}+1)\times(n_{\bullet 1}+1)\times n_{10}\times n_{01}}{(n_{11}+1)^2\times(n_{11}+2)},$$

where $n_{11} = n_2$, $n_{10} = n_4$, $n_{01} = n_6$, $n_{1\bullet} = n_{11}+n_{10}$, and $n_{\bullet 1} = n_{11}+n_{01}$.

While neither standard estimator is efficient under the anchor stream design, that design validates a maximum likelihood estimator (MLE) under a general multinomial model for the population-level cell counts that was originally proposed for two-stream CRC sensitivity analyses (7; 20, 21):

$$\hat{N}_\psi = n_{11}+n_{10}+n_{01}/\psi \quad, \quad (2)$$

where $\psi = p_{2|\bar{1}}$ is the probability of a case being identified in Stream 2 given he/she is not identified in Stream 1. In standard CRC the parameter $p_{2|\bar{1}}$ is never known, but $\hat{N}_{Chap}$ estimates N based on the LP assumption that dictates equality of $p_{2|\bar{1}}$ and the analogous estimable parameter $p_{2|1}$. The key to the potential precision offered by $\hat{N}_\psi$ in (2) is that the anchor stream design (with $N_{tot}$ known) ensures not only the LP conditions but the equality of $p_{2|\bar{1}}$ with the selection probability ($\psi$) into the Stream 2 random sample, which can be treated as known and equal to the sampling rate ($\psi = n_{RS}/N_{tot}$). An estimator of the variance characterizing the estimator in eqn. (2) is as follows (20, 21):

$$\text{Vâr}(\hat{N}_\psi) = n_{01}(1-\psi)/\psi^2 \qquad (3)$$

Under the anchor stream design, the estimator in eqn. (2) may not be fully efficient, and the variance in eqn. (3) can be conservative as the true disease prevalence (p) increases. Empirically (see **Simulation Studies**), we find eqn. (2) to be efficient and eqn. (3) reliable when prevalence is relatively small (e.g., $p \leq 0.2$).

Variants of the estimator in eqn. (2) were found to be much more precise than classical CRC estimators (7). In particular, we recommend the MLE based on the full multinomial model (conditional on the total population size $N_{tot}$) for the cell counts in Table 1:

$$\hat{N}_{\hat{\psi}*} = n_2 + n_4 + n_6 \left( \frac{n_5 + n_6 + n_7}{n_5 + n_6} \right) \qquad (4)$$

An approximate variance to accompany $\hat{N}_{\hat{\psi}*}$ was provided (7):

$$\text{Vâr}(\hat{N}_{\hat{\psi}*}) = \left[ \text{Vâr}^{-1}(\hat{N}_{RS}) + \text{Vâr}^{-1}(\hat{N}_{LP}) \right]^{-1}, \qquad (5)$$

where $\text{Vâr}(\hat{N}_{LP}) = \frac{(n_{11}+n_{10}) \times (n_{11}+n_{01}) \times n_{10} \times n_{01}}{n_{11}^3}$ is the estimated variance of the classical Lincoln-Petersen estimator (15) and we calculate $\text{Vâr}(\hat{N}_{LP})$ after replacing any 0 observed cell ($n_{11}$, $n_{10}$, or $n_{01}$) by 0.5.

Our evaluations suggest $\hat{N}_{\hat{\psi}*}$ is essentially fully efficient among estimators under the anchor stream design that assume nothing about the mechanism behind selection into Stream 1. An efficient anchor stream-based estimator of current disease prevalence (p) follows immediately:

$$\hat{p}_{\hat{\psi}*} = \hat{N}_{\hat{\psi}*} / N_{tot} \qquad (6)$$

While a Wald-type confidence interval (CI) for N or p can be based on the variance in eqn. (3) when p is small, we prefer an adapted version of a Dirichet-multinomial-based Bayesian credible interval that leverages added precision when p is moderate to large (see *Interval Estimation*). In *Appendix A*, we update a similar approach (7), in order to target favorable coverage across a wide range of disease prevalences.

### *Estimating General Means Under the Anchor Stream Design*

Suppose now the goal is to estimate the mean of a variable X in the overall closed population. If X is binary, the estimator in eqn. (4) applies. However, X might represent something more general (e.g., a continuous biomarker level). In this case, we find that the anchor stream design enables a direct standardization-type estimator of the mean of X. In general, the estimator would be:

$$\hat{\mu}_x = \bar{x}_{11}\hat{p}_{11} + \bar{x}_{10}\hat{p}_{10} + \bar{x}_{01}\hat{p}_{01} + \bar{x}_{00}\hat{p}_{00} ,$$

where the (i, j) subscripts indicate capture status (yes/no) by Streams 1 and 2, $\hat{p}_{ij}$ is the observed proportion of the $N_{tot}$ subjects with capture history (i, j), and $\bar{x}_{ij}$ is the sample mean of X among those with capture history (i, j). Typically the above estimator could not be used, because there would be no defensible way to obtain the necessary estimates $\hat{p}_{00}$ and $\bar{x}_{00}$ characterizing the unseen (0, 0) population. However, the anchor stream design overcomes this and allows standardization based only on whether individuals were identified in Stream 1, i.e.,

$$\hat{\mu}_x = \bar{x}_{1\bullet}\hat{p}_{1\bullet} + \bar{x}_{0\bullet}\hat{p}_{0\bullet} , \qquad (7)$$

where $\bar{x}_{1\bullet}$ and $\bar{x}_{0\bullet}$ are estimates of the mean of X among those captured and not captured in Stream 1, $\hat{p}_{1\bullet}$ is the proportion of the $N_{tot}$ population members observed in Stream 1, and $\hat{p}_{0\bullet} = 1 - \hat{p}_{1\bullet}$. In eqn. (7), $\bar{x}_{1\bullet}$ is the sample mean of X among those identified in Stream 1. The crucial parameter identified by design is the mean among those <u>not</u> included in Stream 1, estimable as the sample mean of X among those identified in Stream 2 but not in Stream 1 (i.e., $\bar{x}_{0\bullet} = \bar{x}_{01}$).

If X is case status (or any other binary variable), then eqn. (7) is exactly equivalent to the prevalence estimator $\hat{p}_{\hat{\psi}*} = \hat{N}_{\hat{\psi}*}/N_{tot}$ in eqn. (6). Eqn. (7), however, has the unique advantage of applying to estimating the overall mean of any variable X. To estimate the mean of X (e.g., a continuous biomarker level) exclusively for <u>cases</u> in the target population, we assume case status and X are observed for those sampled in Streams 1 and 2 and use a modified version of eqn. (7):

$$\hat{\mu}_{x,cases} = \bar{x}_{1\bullet,cases}\hat{p}_{1\bullet,cases} + \bar{x}_{0\bullet,cases}\hat{p}_{0\bullet,cases}, \qquad (8)$$

where $\bar{x}_{1\bullet,cases}$ is the sample mean of X among cases identified in Stream 1, $\bar{x}_{0\bullet,cases} = \bar{x}_{01,cases}$ is the sample mean of X among cases identified in Stream 2 but not 1, $\hat{p}_{1\bullet,cases} = n_{1\bullet}/\hat{N}_{\psi*}$, $n_{1\bullet}$ is the number of cases identified in Stream 1, and $\hat{p}_{0\bullet,cases} = 1 - \hat{p}_{1\bullet,cases}$. To estimate the mean of X for <u>non-cases</u>, eqn. (8) applies upon changing the label "cases" to "non-cases". This allows biomarker evaluation via valid estimation of the difference in the mean for cases vs. non-cases.

*Interval Estimation*

Using only the Stream 2 random sample, one can calculate a normal theory-based CI using the FPC-adjusted standard error corresponding to expression (1). However, such Wald-type CIs often behave poorly in small sample settings (22). Bayesian credible intervals for binomial proportions based on a weakly informative Jeffrey's beta prior typically provide more reliable frequentist coverage when sample size is small and/or prevalence is extreme (23). In *Appendix A*, we incorporate the FPC adjustment into this Bayesian credible interval paradigm; we use this approach to evaluate CIs based on Stream 2 alone (see *Simulation Studies*).

When X is binary, the case count estimators in expressions (2) and (4) and their associated prevalence estimators apply. We evaluate these estimators and assess the proposed anchor stream-based credible interval approach (see Appendix A) in *Simulation Studies*. When estimating more general means (e.g., when X is non-binary), we propose conducting inference via bootstrap percentile intervals (24). Bootstrap samples are drawn based on the observed data from those captured at least once in Streams 1 and 2. Details of the calculations appear in *Appendix B*.

*Sample Size Planning*

For planning the size of the anchor stream random sample, one could use the variance in eqn. (1) together with an assumed prevalence and desired precision level. However, this applies only to the estimator $\hat{N}_{RS}$ and will often be highly conservative under the proposed design with $\hat{N}_{\hat{\psi}*}$ in eqn. (4) as the recommended estimator. For more refined guidance, we offer the following approximation for the required Stream 2 sampling rate ($\psi$):

$$\psi = p(1-\phi_1)\left[ N_{tot}^2 \sigma_p^2 + p(1-\phi_1) \right]^{-1} \qquad (9)$$

In expression (9), $\sigma_p$ is the desired standard error sought in conjunction with the prevalence estimator $\hat{p}_{\hat{\psi}*}$ in eqn. (6), p is the assumed prevalence, and $\phi_1$ the assumed proportion of diseased cases to be identified in Stream 1 (this could be very different from the overall sampling rate into Stream 1). We propose eqn. (9) for use in relatively low prevalence settings (e.g., $p \leq 0.2$), where we expect the estimator in eqn. (2) and its variance in eqn. (3) to be efficient and reliable. As prevalence increases, eqn. (9) tends to overestimate the Stream 2 sampling rate.

**Results**

We conducted two series of simulations to evaluate estimators under the anchor stream design, studying closed-population estimation and inference with respect to case counts (Series 1) and a continuous (e.g., biomarker) mean (Series 2). In *Appendix C*, we provide a worked example based on actual cell counts ($n_1$—$n_7$; Table 1), along with a GitHub link with R code that expands further on the example.

*Simulation Studies*

For Series 1 simulations, we examined a wide range of scenarios with respect to population size ($N_{tot}$ = 50, 100, 250, 500, 1000), disease prevalence (p = 0.05, 0.1, 0.2, 0.5), and the Stream 2 sampling rate ($\psi$ = 0.1, 0.2, 0.4, 0.5). For Series 2, we examined multiple scenarios by varying $N_{tot}$ (250, 500, 1000) and disease prevalence (p = 0.1, 0.2, 0.5), with the Stream 2 sampling rate fixed at $\psi$ = 0.2. For simulated diseased and non-diseased individuals, symptom status was randomly generated as binomial with a 50% and 10% success rate, respectively. To mimic preferential self-selection for voluntary testing, participation in Stream 1 was then generated with 90% and 20% success rates for those with and without symptoms, respectively. We conducted 10,000 simulations under each scenario. Adjusted credible intervals (Appendix A) for Series 1 inference were based on 10,000 Dirichlet posterior draws, while bootstrap-based inference for Series 2 (Appendix B) used 1,000 bootstrap replicates per simulation.

**Table 2** summarizes Series 1 simulations with $N_{tot}$ = 500 and small prevalence (p = 0.05, 0.1). Notably, the anchor stream design-based estimators $\hat{N}_\psi$ and $\hat{N}_{\hat{\psi}*}$ are far more precise than $\hat{N}_{RS}$ based solely on the Stream 2 random sample. For inference associated with $\hat{N}_{RS}$, we assess an FPC-adjusted Wald-type CI together with our proposed Jeffrey's prior-based FPC-adjusted credible interval (Appendix A). This Wald-type CI is the only interval in Table 2 for which we did not incorporate the observed number of cases ($n_2+n_4+n_6$) in Table 1 as a minimum for the lower bound; this was in order to assess interval widths if only Stream 2 data had been collected. The narrowness of the credible intervals relative to the Wald CIs accompanying $\hat{N}_{RS}$ stems primarily from our incorporation of the case number bound enabled by the data in Table 1. Regarding coverage, the Jeffrey's-based interval outperforms the Wald CI in all but one case in Table 2.

With the anchor stream estimators $\hat{N}_\psi$ and $\hat{N}_{\hat{\psi}*}$ in Table 2, we similarly assess both Wald-type CIs and adjusted Bayesian credible intervals (Appendix A). The results demonstrate favorable coverage of the proposed credible interval, which outperforms Wald-type alternatives and yields narrower intervals than those based on Stream 2 alone. The credible intervals presented with $\hat{N}_\psi$ and $\hat{N}_{\hat{\psi}*}$ in Table 2 are identical, as the low prevalence dictates that the unadjusted credible interval (Appendix A) was applied in both cases. The estimators $\hat{N}_\psi$ and $\hat{N}_{\hat{\psi}*}$ also yield nearly identical empirical SDs in Table 2.

**Table 3** summarizes Series 1 simulations with $N_{tot} = 500$, but larger disease prevalence (p = 0.2, 0.5), to demonstrate the benefits of fine-tuning inferences (Appendix A). Here, there is little difference in coverage between the FPC-adjusted Bayesian and Wald intervals in conjunction with $\hat{N}_{RS}$. The key precision gains of the anchor stream estimators relative to $\hat{N}_{RS}$ remain apparent, and we begin to see improved efficiency offered by $\hat{N}_{\hat{\psi}*}$ relative to $\hat{N}_\psi$. As prevalence increases (p=0.5), both this trend and a tendency toward conservative estimated SEs accompanying $\hat{N}_\psi$ [eqn. (3)] become evident. The adjusted Bayesian credible interval (Appendix A) associated with $\hat{N}_{\hat{\psi}*}$ is now noticeably narrowest, and continues to yield near-nominal coverage.

Improvements in interval width enabled by the anchor stream design occurred even though we incorporated the observed number of cases in Table 1 as a minimal lower bound for credible intervals based solely on the random sample. Taken together, Tables 2 and 3 favor the adjusted Bayesian credible interval proposed in conjunction with $\hat{N}_{\hat{\psi}*}$ (Appendix A), given its reliable performance and favorable width across the full range of conditions. For an expanded set of simulation conditions examining other $N_{tot}$ values (50, 100, 250, 1000, 5000), we refer to ***Online Appendix*** (Tables S1—S6).

For Series 2 simulations, we generated a continuous random variable X representing a biomarker. To mimic heterogeneity in its distribution as might be expected in practice, we simulated X via a mixture of distributions with mean and variance changing with symptom and disease status. Specifically, X was drawn as a normal random variate within each of 4 strata defined by the symptom and disease indicators, with mean and standard deviation ($\mu$, $\sigma$) combinations of (10, 0.75), (5, 0.5), (2.5, 1.2), and (1, 1.5) for those in (symptom, disease) strata (1, 1), (0, 1), (1, 0), and (0, 0), respectively. The goal of the Series 2 studies is to show how the anchor stream design allows leveraging information from an arbitrarily non-representative sample (Stream 1) to gain precision when estimating the overall mean ($\mu_X$) of a continuous biomarker X. We also demonstrate how the design simultaneously unlocks precision gains when estimating the mean of X separately for cases and non-cases. The simulation conditions here dictate a true overall $\mu_X$ that varies with the assumed disease prevalence (p). Specifically, $\mu_X$ = 1.785, 2.42, and 4.325 when p = 0.1, 0.2, and 0.5, respectively. In contrast, the means of X among cases and non-cases were fixed ($\mu_{X,cases} = 7.5$, $\mu_{X,noncases} = 1.15$) across each set of conditions, yielding the same mean difference ($\mu_{diff} = 6.35$).

**Table 4** summarizes Series 2 simulations with $N_{tot}$ = 500 and $\psi$ = 0.2. Note that the estimator $\bar{x}_{1\bullet}$ is severely biased upward due to non-representative sampling in Stream 1. We do not report estimated SEs or CI coverage in conjunction with $\bar{x}_{1\bullet}$, as standard FPC adjustments do not correctly account for non-random finite population sampling. The estimator $\bar{x}_{2\bullet}$ based only on Stream 2 is unbiased and the SE in eqn. (1) is valid, as expected. We assess the corresponding Wald-type CI, which was nearly identical to an FPC-adjusted percentile interval based on bootstrap replicates of $\bar{x}_{2\bullet}$ (results not shown). A key message from Table 4 is that CI coverages in conjunction with $\bar{x}_{2\bullet}$ are essentially matched by the bootstrap percentile CIs (Appendix B) accompanying the proposed anchor stream estimator $\hat{\mu}_x$ in eqn. (7). However, by incorporating additional information from the non-representative Stream 1 sample, $\hat{\mu}_x$ achieves marked precision improvements and narrower CIs.

The lower part of Table 4 summarizes estimates of the mean of X among cases and non-cases. Again, the estimator $\bar{x}_{1\bullet,cases}$ is seriously biased because Stream 1 identifies a non-random sample of cases. With no FPC effect upon restricting attention to cases, a standard bootstrap-based SE is valid here; however, the bias in $\bar{x}_{1\bullet,cases}$ yields CIs with near-null coverage. In contrast, the Stream 2 estimator $\bar{x}_{2\bullet,cases}$ and the proposed case-only anchor stream estimator $\hat{\mu}_{x,cases}$ are both reliable, as are bootstrap-based SEs and percentile CIs without FPC adjustments (Appendix B). Again, $\hat{\mu}_{x,cases}$ is far more precise than $\bar{x}_{2\bullet,cases}$, with narrower CIs. The same conclusion applies when comparing mean estimators for non-cases ($\bar{x}_{2\bullet,noncases}$ vs. $\hat{\mu}_{x,noncases}$), as well estimators of the mean difference for cases relative to non-cases ($\bar{x}_{2\bullet,diff}$ vs. $\hat{\mu}_{x,diff}$). For qualitatively similar Series 2 simulation results across a range of $N_{tot}$ values (250, 1000, 5000), see ***Online Appendix*** (Tables S7—S8).

**Discussion**

We set out to promote an "anchor stream" study design for disease monitoring in closed communities. The key appeal of this design is that it allows use of data from an existing surveillance stream (Stream 1, e.g., based on voluntary testing), which could be arbitrarily non-representative of the closed target population, toward a reliable estimate of the case count. It does this by augmenting Stream 1 with a random sample (Stream 2), implemented according to the prescribed design. The Stream 2 sample could be small, saving resources by drawing additional precision from the Stream 1 data. Part of our purpose has been to evaluate anchor stream-based point estimators of the case count across a wide range of true prevalences, and to refine corresponding inferential procedures (Appendix A). We refer to (7) for guidance on incorporating multiple non-anchor streams and for comparisons against the classical CRC estimator of Chapman (14), which was decidedly less precise than all estimators summarized in Tables 2-3 and thus not further investigated here. Our second purpose has been to exploit an alternative representation [eqn. (7)] of the case count estimator in eqn. (4), demonstrating that the anchor stream design accommodates estimation of general means (e.g., of biomarkers) in two-stream surveillance.

Regarding the apparently novel concept of estimating general means in a CRC-type setting, our work suggests this is also possible without knowing the total population size and without requiring a list of closed population members for random sampling. That is, the estimation of E(X), e.g., for a continuous variable, could be accomplished using a variant on eqn. (7) whereby the population size ($N_{tot}$) is first estimated by *assuming* that Stream 2 achieves a random sample independently of the non-representative Stream 1. The analogue to eqn. (7) would utilize the estimated $N_{tot}$ (e.g., via Chapman's estimator) in the denominator to calculate $\hat{p}_{1\bullet}$, in place of the known $N_{tot}$ as prescribed under the anchor stream design. Investigating such an estimator may be of interest, though it might be much less efficient than the anchor stream-based version and certainly much less robust due to assuming the necessary conditions rather than ensuring them by design. Even with an anchor stream, we note that estimating E(X) for a continuous variable can be problematic in small population or small prevalence scenarios when there are very few individuals available to produce the estimate $\bar{x}_{0\bullet} = \bar{x}_{01}$ in eqn. (7) or $\bar{x}_{0\bullet,cases} = \bar{x}_{01,cases}$ in eqn. (8).

As noted in *Simulation Studies*, our control of the self-selection process into Stream 1 for simulation purposes enables an "oracle-type" direct standardization estimate of prevalence based only on that non-representative stream. That is, an all-knowing analyst would have an estimator of the form $\hat{\mu}_x = \overline{x}_{1,sympt=1}\hat{p}_{sympt=1} + \overline{x}_{1,sympt=0}\hat{p}_{sympt=0}$, where $\overline{x}_{1,sympt=1}$ is the sample mean of X among those selected in Stream 1 with symptoms, $\hat{p}_{sympt=1}$ is the proportion of the total population with symptoms, and the other terms are defined analogously. This could be either more or less precise than the estimator in eqn. (7), but the latter is robust to any Stream 1 selection process when the anchor stream design is applied. The "oracle", however, is never directly available in observational settings and must be replaced, e.g., by a version based on strata defined by observable covariates believed to be associated with X and to explain the non-representative sampling (see (25) for a recent example). While causal inferential methods could be leveraged in efforts to improve robustness of such estimators, variables like symptom status (which could be key drivers of voluntary testing in infectious disease monitoring) are constantly changing from person to person and unlikely to be measurable across the full target population. Thus, attempts to approach the oracle could fall woefully short of accounting for non-representativeness.

Through ongoing work, we hope to explore other versions of the anchor stream design that could be applied to surveillance and to extend the analytical methods to make them more broadly applicable. We envision electronic health record (EHR)-based uses of this design for studying prevalent characteristics of clinical cohorts, when the initial trigger to assess them on a given individual occurs only by indication. We also view the extension to allow for misclassified case assessment in one or both surveillance streams as a development that could have useful applications toward surveillance of chronic and infectious diseases. Here, we have assumed the use of a diagnostic testing method with sufficient accuracy to be viewed as an arguable gold standard (e.g., PCR-based testing in the context of COVID-19) for disease monitoring.

## ACKNOWLEDGEMENTS

We thank Drs. Kevin Ward, Tim Lash, Aaron Siegler and Sarita Shah for enlightening motivation and discussion. Partial support was provided by the National Institute of Health-funded Emory Center for AIDS Research (P30AI050409; Del Rio PI), the National Center for Advancing Translational Sciences of the National Institutes of Health (UL1TR002378; Taylor PI), the Eunice Kennedy Shriver National Institute of Child Health and Human Development (R01HD092508; Waller PI), and by the National Institutes of Health/National Cancer Institute-funded Cancer Recurrence Information and Surveillance Program (CRISP) study (1 R01 CA208367-01; Ward/Lash MPIs). The content is solely the responsibility of the authors.

**Table 1. Observations and Cell Counts Based on the Anchor Stream Design**

| Cell Count | Observation Type |
|---|---|
| $n_1$ | Sampled in Both Streams, Tested Negative |
| $n_2$ | Sampled in Both Streams, Tested Positive |
| $n_3$ | Sampled in Stream 1 but Not 2, Tested Negative |
| $n_4$ | Sampled in Stream 1 but Not 2, Tested Positive |
| $n_5$ | Sampled in Stream 2 but Not 1, Tested Negative |
| $n_6$ | Sampled in Stream 2 but Not 1, Tested Positive |
| $n_7$ | Not Sampled in Either Stream |

**Table 2. Simulations Evaluating Case Count Estimates with Low Prevalence and $N_{tot} = 500$**

| | True Disease Prevalence (p) = 0.05 | | | | | |
|---|---|---|---|---|---|---|
| | $\psi = 0.1$ | | $\psi = 0.2$ | | $\psi = 0.5$ | |
| Estimator | Mean (SD) [avg. SE] | CI coverage [avg. width] | Mean (SD) [avg. SE] | CI coverage [avg. width] | Mean (SD) [avg. SE] | CI coverage [avg. width] |
| $\hat{N}_{RS}$ [a] | 25.0 (14.8) [14.2] | 92.8%, **89.8%** [51.0], **[49.7]** | 24.9 (9.8) [9.5] | 90.0%, **95.8%** [37.1], **[32.8]** | 25.0 (4.9) [4.9] | 94.3%, **96.0%** [19.0], **[16.5]** |
| $\hat{N}_{\psi}$ [b] | 24.9 (9.9) [10.2] | 95.9%, **98.5%** [30.0], **[39.2]** | 24.9 (6.6) [6.4] | 88.4%, **94.3%** [21.5], **[25.1]** | 25.0 (3.3) [3.3] | 91.6%, **95.3%** [12.0], **[12.5]** |
| $\hat{N}_{\hat{\psi}*}$ [c] | 25.0 (9.9) [9.0] | 84.7%, **98.5%** [27.1], **[39.2]** | 24.9 (6.6) [6.1] | 85.8%, **94.3%** [20.2], **[25.1]** | 25.0 (3.3) [3.1] | 90.0%, **95.3%** [11.3], **[12.5]** |
| | True Disease Prevalence (p) = 0.1 | | | | | |
| $\hat{N}_{RS}$ [a] | 49.8 (20.0) [19.7] | 89.8%, **95.5%** [76.5], **[67.5]** | 49.7 (13.4) [13.3] | 94.9%, **96.0%** [52.1], **[46.5]** | 49.9 (6.7) [6.7] | 93.7%, **94.9%** [26.3], **[24.1]** |
| $\hat{N}_{\psi}$ [b] | 50.0 (13.9) [13.7] | 89.9%, **95.0%** [46.8], **[55.0]** | 50.0 (9.4) [9.2] | 91.1%, **95.4%** [34.6], **[36.4]** | 50.0 (4.6) [4.7] | 94.3%, **95.5%** [18.4], **[18.4]** |
| $\hat{N}_{\hat{\psi}*}$ [c] | 50.0 (13.9) [12.8] | 85.8%, **95.0%** [43.7], **[55.0]** | 49.9 (9.3) [8.8] | 89.5%, **95.4%** [32.8], **[36.4]** | 50.0 (4.6) [4.5] | 93.0%, **95.5%** [17.6], **[18.4]** |

a  SE based on (1); Wald-based CIs are evaluated (non-bold) along with a proposed FPC-adjusted Jeffreys prior-based credible interval (**bold**; see Appendix A)

b  SE based on (3); Wald-based CIs are evaluated (non-bold) along with a proposed unadjusted Dirichlet-multinomial-based credible interval (**bold**; see Appendix A)

c  SE based on (5); Wald-based CIs are evaluated (non-bold) along with a proposed adjusted Dirichlet-multinomial-based credible interval (**bold**; see Appendix A)

**Table 3. Simulations Evaluating Case Count Estimates with Higher Prevalence and $N_{tot} = 500$**

| | True Disease Prevalence (p) = 0.2 | | | | | |
|---|---|---|---|---|---|---|
| | $\psi = 0.1$ | | $\psi = 0.2$ | | $\psi = 0.5$ | |
| Estimator | Mean (SD) [avg. SE] | CI coverage [avg. width] | Mean (SD) [avg. SE] | CI coverage [avg. width] | Mean (SD) [avg. SE] | CI coverage [avg. width] |
| $\hat{N}_{RS}$ [a] | 99.5 (26.7) [26.7] | 95.2%, **96.1%** [104.5], **[93.2]** | 99.8 (17.9) [17.8] | 94.9%, **94.9%** [70.0], **[65.9]** | 99.9 (8.9) [8.9] | 95.5%, **94.4%** [35.1], **[34.2]** |
| $\hat{N}_{\psi}$ [b] | 100.0 (19.0) [19.6] | 92.2%, **95.8%** [74.3], **[78.2]** | 100.0 (12.7) [13.3] | 93.8%, **96.2%** [51.9], **[52.0]** | 100.0 (6.4) [6.7] | 95.4%, **96.0%** [26.2], **[26.2]** |
| $\hat{N}_{\hat{\psi}*}$ [c] | 99.9 (18.7) [18.3] | 90.0%, **95.4%** [68.9], **[77.1]** | 100.0 (12.5) [12.5] | 92.2%, 95.9% [48.6], [51.4] | 100.0 (6.3) [6.3] | 94.2%, **95.8%** [24.7], **[25.7]** |
| | True Disease Prevalence (p) = 0.5 | | | | | |
| $\hat{N}_{RS}$ [a] | 249.4 (33.4) [33.6] | 95.0%, **95.0%** [131.6], **[126.9]** | 249.7 (22.3) [22.4] | 94.5%, **94.5%** [87.7], **[86.6]** | 250.0 (11.2) [11.2] | 94.2%, **94.2%** [43.9], **[43.7]** |
| $\hat{N}_{\psi}$ [b] | 250.2 (28.5) [31.6] | 95.5%, **97.4%** [123.8], **[124.5]** | 249.9 (18.7) [21.1] | 96.4%, **97.2%** [82.8], **[82.8]** | 250.2 (9.4) [10.6] | 97.1%, **97.2%** [41.6], **[41.5]** |
| $\hat{N}_{\hat{\psi}*}$ [c] | 250.0 (25.3) [26.0] | 93.7%, **95.7%** [102.1], **[110.2]** | 249.9 (16.7) [17.5] | 94.9%, **95.9%** [68.5], **[72.5]** | 250.1 (8.5) [8.8] | 95.4%, **95.7%** [34.4], **[36.2]** |

a SE based on (1); Wald-based CIs are evaluated (non-bold) along with a proposed FPC-adjusted Jeffreys prior-based credible interval (**bold**; see Appendix A)

b SE based on (3); Wald-based CIs are evaluated (non-bold) along with a proposed unadjusted Dirichlet-multinomial-based credible interval (**bold**; see Appendix A)

c SE based on (5); Wald-based CIs are evaluated (non-bold) along with a proposed adjusted Dirichlet-multinomial-based credible interval (**bold**; see Appendix A)

**Table 4. Simulations Evaluating Mean Estimates for Continuous X with $N_{tot}$ = 500 and $\psi$ =0.2**

| | Overall Mean Estimators | | | | | |
|---|---|---|---|---|---|---|
| | **p = 0.1, True $\mu_x$ =1.785** | | **p = 0.2, True $\mu_x$ =2.42** | | **p = 0.5, True $\mu_x$ =4.325** | |
| Estimator | Mean (SD) [avg. SE] | CI coverage [avg. width] | Mean (SD) [avg. SE] | CI coverage [avg. width] | Mean (SD) [avg. SE] | CI coverage [avg. width] |
| $\bar{x}_{1\bullet}$ [a] | 2.904 (0.22) [--] | -- | 4.064 (0.25) [--] | -- | 6.596 (0.24) [--] | -- |
| $\bar{x}_{2\bullet}$ [b] | 1.785 (0.24) [0.23] | 93.1% [0.89] | 2.422 (0.29) [0.28] | 93.8% [1.09] | 4.326 (0.35) [0.34] | 94.1% [1.35] |
| $\hat{\mu}_x$ [c] | 1.785 (0.16) [0.15] | 93.0% [0.59] | 2.420 (0.18) [0.17] | 93.5% [0.68] | 4.323 (0.21) [0.20] | 93.8% [0.80] |
| | Mean Estimators for Cases | | | | | |
| | **p = 0.1, True $\mu_{x,cases}$ =7.5** | | **p = 0.2, True $\mu_{x,cases}$ =7.5** | | **p = 0.5, True $\mu_{x,cases}$ =7.5** | |
| $\bar{x}_{1\bullet,cases}$ [d] | 9.090 (0.39) [0.39] | 5.6% [1.52] | 9.087 (0.28) [0.28] | 0.1% [1.08] | 9.088 (0.18) [0.18] | 0.0% [0.68] |
| $\bar{x}_{2\bullet,cases}$ [e] | 7.493 (0.85) [0.84] | 93.0% [3.28] | 7.492 (0.59) [0.58] | 94.6% [2.27] | 7.505 (0.37) [0.37] | 94.4% [1.43] |
| $\hat{\mu}_{x,cases}$ [f] | 7.560 (0.60) [0.55] | 94.0% [2.13] | 7.524 (0.42) [0.41] | 94.2% [1.59] | 7.511 (0.26) [0.24] | 93.2% [0.95] |
| | Mean Estimators for Non-Cases | | | | | |
| | **p=0.1, True $\mu_{x,noncases}$ =1.15** | | **p=0.2, True $\mu_{x,noncases}$ =1.15** | | **p=0.5, True $\mu_{x,noncases}$ =1.15** | |
| $\bar{x}_{1\bullet,noncases}$ | 1.499 (0.14) [0.14] | 31.7% [0.56] | 1.501 (0.15) [0.15] | 35.8% [0.59] | 1.506 (0.19) [0.19] | 53.3% [0.75] |
| $\bar{x}_{2\bullet,noncases}$ | 1.145 (0.16) [0.16] | 94.0% [0.63] | 1.153 (0.18) [0.17] | 94.0% [0.67] | 1.146 (0.22) [0.22] | 94.7% [0.85] |
| $\hat{\mu}_{x,noncases}$ | 1.147 (0.14) [0.14] | 93.6% [0.55] | 1.152 (0.15) [0.15] | 94.0% [0.58] | 1.144 (0.19) [0.19] | 94.6% [0.74] |
| | Estimators for Case vs. Non-Case Mean Difference | | | | | |
| | **p=0.1, True $\mu_{diff}$ =6.35** | | **p=0.2, True $\mu_{diff}$ =6.35** | | **p=0.5, True $\mu_{diff}$ =6.35** | |
| $\bar{x}_{1\bullet,diff}$ | 7.591 (0.41) [0.42] | 21.9% [1.62] | 7.586 (0.32) [0.32] | 4.7% [1.23] | 7.583 (0.26) [0.26] | 0.4% [1.01] |
| $\bar{x}_{2\bullet,diff}$ | 6.337 (0.86) [0.85] | 93.3% [3.36] | 6.320 (0.61) [0.61] | 94.4% [2.37] | 6.344 (0.43) [0.43] | 94.6% [1.67] |
| $\hat{\mu}_{x,diff}$ | 6.385 (0.62) [0.57] | 93.2% [2.23] | 6.359 (0.45) [0.44] | 94.5% [1.72] | 6.360 (0.33) [0.31] | 93.7% [1.23] |

a Estimated mean for individuals sampled in Stream 1; SE and CIs not reported

b Estimated mean for individuals sampled in Stream 2; SE incorporates FPC adjustment with Wald-type CIs; SE of $\bar{x}_{2\bullet}$ equals sample standard deviation of X among those sampled in Stream 2 divided by square root of the number of individuals sampled in Stream 2

c SE based on bootstrap with percentile CIs incorporating FPC adjustments (see Appendix B)

d Estimated mean among cases sampled in Stream 1; SE based on bootstrap with percentile CIs

e Estimated mean among cases sampled in Stream 2; SE based on bootstrap with percentile CIs

f SE based on bootstrap with percentile CIs (see Appendix B)

# Appendix A: Adjusted Bayesian Credible Intervals for Case Counts

*Inference Based on Stream 2 Only*

First, we focus on inference about the true prevalence or case count based only on the anchor Stream 2 random sample. We seek to leverage potential improvements in frequentist coverage properties made possible by beta distribution-based posterior credible intervals relative to standard Wald-type confidence intervals (CIs), while accounting for the associated finite population correction (FPC). Relevant literature in the absence of the FPC is extensive (see for example Agresti and Coull 1998; Brown, Cai and DasGupta 2001; Lyles, Weiss and Waller 2020), while guidance related to finite population sampling is scarce and tends to be presented from a theoretical perspective (e.g., Meeden and Vardeman 1991). Here, we offer a practical and effective approach, based on a simple scale and shift adjustment to the a typical posterior credible interval for the prevalence based on a Jeffreys Beta(0.5, 0.5) prior distribution. Referring to eqn. (1) for notation, the resulting conjugate posterior is well known to be Beta($n_{RS}^+ + 0.5, n_{RS} - n_{RS}^+ + 0.5$), so that the corresponding 95% credible interval would be defined using the 2.5$^{th}$ and 97.5$^{th}$ percentiles of that distribution (e.g., Brown, Cai and DasGupta 2001). To account for the FPC as defined in eqn. (1), our approach is effectively to first scale each Beta posterior draw by multiplying it by $a = \sqrt{FPC}$, and then add to each scaled draw the term $b = \hat{p}_{RS}(1-a)$. These measures adjust the posterior distribution to have variance equal to FPC times the original variance, and mean equal to $\hat{p}_{RS}$. We then take the 2.5$^{th}$ and 97.5$^{th}$ percentiles to be an FPC-adjusted version of the Jeffreys interval. Because the original posterior is easily accessed in most statistical software packages, however, one need not actually take any posterior draws directly. Rather, the proposed FPC-adjusted interval is defined by the following two readily accessible percentiles:

$$\left[ Beta(0.025; n_{RS}^+ + 0.5, n_{RS} - n_{RS}^+ + 0.5), \ Beta(0.975; n_{RS}^+ + 0.5, n_{RS} - n_{RS}^+ + 0.5) \right] \quad (A.1)$$

Extensive simulation studies (not all summarized in this report) support the effectiveness of the interval in (A.1) in terms of overall coverage and width, suggesting that it could be considered for general use as a tool for finite population sampling-based inference about binomial proportions. In the simulation studies for the current article, we use the limits in (A.1) multiplied by the total population size ($N_{tot}$) as the "Stream 2 Only" interval estimate for the case count, but we also allow this interval to use the knowledge that the lower limit should be no less than the total number of cases identified in Streams 1 and 2, i.e., no smaller than $n_c = n_{11} + n_{10} + n_{01} = n_2 + n_4 + n_6$ (see Table 1). This gives the "Stream 2 Only" interval its best possible width advantage, for comparison with our proposed approach to make full use of both streams (see next section).

*Inference Incorporating Both Stream 1 and Stream 2*

Empirical studies examined in this article evaluate the estimator $\hat{N}_\psi$ in eqn. (2) along with a Wald-type CI utilizing the variance given in (3), i.e., $\hat{N}_\psi \pm 1.96 \text{SE}(\hat{N}_\psi)$. While this interval is seen to perform well in small prevalence settings, we seek an approach that can be recommended across a broad range of conditions. For this purpose, we propose an update of an adjusted Bayesian credible interval originally proposed in Lyles et al. (2022), based on a weakly informative Dirichlet prior on cell probabilities. As in that reference, we begin with a Dirichlet(½, ½, ½) prior for three multinomial proportions $\mathbf{p}^* = (p_{11}^*, p_{10}^*, p_{01}^*)'$, where the $p_{ij}^*$'s represent the probabilities that Stream 1 capture status=i and Stream 2 capture status=j, conditional on being captured by at least one of the two streams. The well-known conjugate posterior distribution is also Dirichlet, i.e.,

$$\mathbf{p}^* | \mathbf{n} \sim Dirichlet(n_{11} + \tfrac{1}{2}, n_{10} + \tfrac{1}{2}, n_{01} + \tfrac{1}{2}), \qquad (A.2)$$

where $\mathbf{n} = (n_{11}, n_{10}, n_{01})'$ is a vector of three primary cell counts (also denoted as $(n_2, n_4, n_6)'$ in Table 1). The next steps make use of the known Stream 2 selection probability ($\psi = n_{RS}/N_{tot}$), along with the general multinomial model underlying the estimator $\hat{N}_\psi$ in eqn. (2). For each of a large sample of draws, $\tilde{\mathbf{p}}_j^*$ (j=1,..., D) using (A.2), we generate the posterior probability of capture in Stream 1 as $\tilde{p}_{1j} = \psi(\tilde{p}_{11}^* + \tilde{p}_{10}^*)\left[\psi(\tilde{p}_{11}^* + \tilde{p}_{10}^*) + \tilde{p}_{01}^*\right]^{-1}$. We then derive the posterior unconditional capture probabilities: $\tilde{p}_{11j} = \tilde{p}_{1j}\tilde{p}_{11}^*(\tilde{p}_{11}^* + \tilde{p}_{10}^*)^{-1}$, $\tilde{p}_{10j} = \tilde{p}_{1j}\tilde{p}_{10}^*(\tilde{p}_{11}^* + \tilde{p}_{10}^*)^{-1}$, and $\tilde{p}_{01j} = \psi(1 - \tilde{p}_{1j})$. The posterior probability of identification in at least one of the two streams then follows as $\tilde{p}_{cj} = \tilde{p}_{1j}(1 - \psi) + \psi$. From this we obtain a draw from the posterior distribution of N conditional on the observed number of unique individuals captured ($n_c = n_{11} + n_{10} + n_{01}$), i.e.,

$$\tilde{N}_{j \mid n_c} = round\left(n_c / \tilde{p}_{cj}\right) \qquad (A.3)$$

To account for uncertainty in the number caught ($n_c$), for the jth draw $\tilde{p}_{cj}$, we generate a new value $n_{cj}$ from the Binomial($\tilde{N}_{j \mid n_c}$, $\tilde{p}_{cj}$) distribution. Posterior draws for the three observed cell counts follow as: $n_{11j} = n_{cj}\tilde{p}_{11j}^*$, $n_{10j} = n_{cj}\tilde{p}_{10j}^*$, and $n_{01j} = n_{cj}\tilde{p}_{01j}^*$. This allows us to compute a posterior draw mimicking the estimand $\hat{N}_\psi$ [eqn. (2)], as follows:

$$\tilde{N}_{\psi,j} = n_{11j} + n_{10j} + n_{01j}/\psi \qquad (A.4)$$

Should any of these draws fall below the original value $n_c$, we set it equal to $n_c$.

At this point, we deviate from the original approach proposed by Lyles et al. (2022), with the goal of producing reliable intervals for N over a broader range of true prevalences. First, we refer to the interval $(LL_{unadj}, UL_{unadj})$ formed by the 2.5$^{th}$ and 97.5$^{th}$ percentiles of the draws in (A.4) as the *Unadjusted* credible interval. For a small estimated prevalence $\hat{p}_{\hat{\psi}*} = \hat{N}_{\hat{\psi}*}/N_{tot}$ (eqn.

6), we recommend this unadjusted credible interval for inference. We suggest 0.2 as a threshold for $\hat{p}_{\hat{\psi}*}$, and used this criterion for simulations described in this article.

When $\hat{p}_{\hat{\psi}*} \geq 0.2$, we propose use of an *Adjusted* credible interval. Our approach involves a scale and shift adjustment based on logic analogous to that leading to (A.1), and is similar to the original proposal from Lyles et al. (2022) with subtle but important differences. In the updated approach, we begin by scaling and re-shifting each draw in (A.4) as follows:

$$\tilde{N}^*_{\psi,j} = a\,\tilde{N}_{\psi,j} + b \quad , \tag{A.5}$$

where $a = \sqrt{\text{Vâr}(\hat{N}_{\hat{\psi}*})/\text{Vâr}(\hat{N}_{\psi})}$, $b = \hat{N}_{\hat{\psi}*}(1-a)$, and the terms $\text{Vâr}(\hat{N}_{\psi})$, $\hat{N}_{\hat{\psi}*}$, and $\text{Vâr}(\hat{N}_{\hat{\psi}*})$ are defined in eqns. (3), (4), and (5), respectively. We denote the 2.5[th] and 97.5[th] sample percentiles of the D draws ($\tilde{N}^*_{\psi,j}; j=1,...,D$) in (A.5) as "$LL_{ab}$" and "$UL_{ab}$", respectively. As a final step, we calculate the lower and upper limits of a Wald-type CI that utilizes the variance of the simple arithmetic average of the estimators $\hat{N}_{RS}$ and $\hat{N}_{Chap}$ (see Section 2). Lyles et al. (2022) provided its estimated variance as $\hat{\sigma}^2_{avg} = \left[\text{Vâr}(\hat{N}_{RS}) + \text{Vâr}(\hat{N}_{Chap})\right]/4$, and we calculate this Wald-type CI as $(LL_{avg}, UL_{avg}) = \hat{N}_{\hat{\psi}*} \pm 1.96\hat{\sigma}_{avg}$. The final adjusted credible interval, recommended and used in our simulation studies whenever $\hat{p}_{\hat{\psi}*} \geq 0.2$, is designed to help mitigate possible anticonservativeness in the interval $(LL_{ab}, UL_{ab})$ while still leveraging potential for improved CI width relative to the unadjusted credible interval when prevalence is moderate to high. The *Adjusted* interval is denoted as $(LL_{adj}, UL_{adj})$, with the limits calculated as follows:

$$LL_{adj} = \min\left(LL_{ab}, \frac{LL_{ab} + LL_{avg}}{2}\right) \quad \text{and} \quad UL_{adj} = \max\left(UL_{ab}, \frac{UL_{ab} + UL_{avg}}{2}\right).$$

**Appendix B: Bootstrap-based Confidence Intervals for General Means**

When estimating general means (e.g., of a continuous biomarker level), either for the overall target population via eqn. (7) or exclusively for cases or non-cases via (8), we propose a bootstrap approach. We begin with the setting of (8), for which we recommend a standard application of a non-parametric bootstrap-based percentile CI (Efron and Tibshirani 1993). Specifically, we start with the observed data records for all individuals identified at least once (i.e., in Stream 1 and/or Stream 2). We randomly draw B bootstrap samples with replacement from this list of individuals (we used B=1,000 in our simulation studies). For the b-th bootstrap sample (b=1,…, B), we first use eqn. (4) to calculate the estimated number of cases (or non-cases) in the target population, i.e.,

$$\hat{N}_{\hat{\psi}^*,b} = n_{2,b} + n_{4,b} + n_{6,b}\left(\frac{n_{5,b} + n_{6,b} + n_{7,b}}{n_{5,b} + n_{6,b}}\right) . \quad (A.6)$$

In the unlikely event that any of the bootstrap replicates in (A.6) should fall below the original number ($n_c$) of individuals identified as a case (or non-case) in either or both streams, it is set to $n_c$. The b-th replicate of the estimand targeting the mean of X among cases then follows eqn. (8), i.e.,

$$\hat{\mu}_{x,cases,b} = \bar{x}_{1\bullet,cases,b}\hat{p}_{1\bullet,cases,b} + \bar{x}_{0\bullet,cases,b}\hat{p}_{0\bullet,cases,b} , \quad (A.7)$$

where $\bar{x}_{1\bullet,cases,b}$ is the mean of X among the cases identified in Stream 1 in the b-th sample, $\bar{x}_{0\bullet,cases,b} = \bar{x}_{01,cases,b}$ is the sample mean of X among the cases identified in Stream 2 but not identified in Stream 1 in the b-th sample, $\hat{p}_{1\bullet,cases,b} = n_{1\bullet,b}/\hat{N}_{\hat{\psi}^*,b}$, and $\hat{p}_{0\bullet,cases,b} = 1 - \hat{p}_{1\bullet,cases,b}$. Note that if a bootstrap sample yields no cases which are identified in Stream 2 but not 1, that sample is not used since $\bar{x}_{01,cases,b}$ is unobtainable. The bootstrap percentile-based CI for E(X) among cases is produced using the 2.5$^{th}$ and 97.5$^{th}$ percentiles of the estimates $\hat{\mu}_{x,cases,b}$ (b=1,…,B).

Our approach to inference about the overall mean E(X) is similar, except that we incorporate FPC adjustments to help mitigate a tendency toward conservative (overly wide) intervals. We base the bootstrap percentile CI on the following replicates (b=1,…, B):

$$\hat{\mu}_{x,b} = \bar{x}^*_{11,b}\hat{p}_{11,b} + \bar{x}^*_{10,b}\hat{p}_{10,b} + \bar{x}^*_{01,b}\hat{p}_{0\bullet,b} , \qquad (A.8)$$

where the asterisks indicate that an FPC is applied. In (A.8), $\hat{p}_{0\bullet,b} = 1 - \hat{p}_{1\bullet,b}$,

$\hat{p}_{1\bullet,b} = (n_{1,b} + n_{2,b} + n_{3,b} + n_{4,b}) / N_{tot}$, $\hat{p}_{11,b} = (n_{1,b} + n_{2,b}) / N_{tot}$, and $\hat{p}_{10,b} = (n_{3,b} + n_{4,b}) / N_{tot}$,

while $\bar{x}^*_{11,b}$, $\bar{x}^*_{10,b}$, and $\bar{x}^*_{01,b}$ are adjusted versions of the sample means ($\bar{x}_{11,b}$, $\bar{x}_{10,b}$, and $\bar{x}_{01,b}$) of X among those in the b-th sample that are identified in both Stream 1 and Stream 2, Stream 1 but not Stream 2, and Stream 2 but not Stream 1, respectively. The adjustments incorporate shift and scale transformations similar in spirit to those utilized in (A.5), based on three FPC factors (FPC$_{11}$, FPC$_{10}$, and FPC$_{01}$) that are calculated based on the original cell counts in Table 1. Specifically,

$$\bar{x}^*_{11,b} = a_{11}\bar{x}_{11,b} + b_{11} , \quad \bar{x}^*_{10,b} = a_{10}\bar{x}_{10,b} + b_{10} , \quad \text{and} \quad \bar{x}^*_{01,b} = a_{01}\bar{x}_{01,b} + b_{01} ,$$

where $a_{11} = \sqrt{FPC_{11}}$, $b_{11} = \bar{x}_{11}(1-a_{11})$, $a_{10} = \sqrt{FPC_{10}}$, $b_{10} = \bar{x}_{10}(1-a_{10})$, $a_{01} = \sqrt{FPC_{01}}$, $b_{01} = \bar{x}_{01}(1-a_{01})$, $\bar{x}_{11}$, $\bar{x}_{10}$, and $\bar{x}_{01}$ are the respective sample means based on the <u>original</u> observed data, and the three FPCs are defined as follows [Cochran 1977; also see eqn. (1)]:

$$FPC_{11} = \frac{n^*_{11}(N^*_{tot11} - n^*_{11})}{N^*_{tot11}(n^*_{11} - 1)} , \quad FPC_{10} = \frac{n^*_{10}(N^*_{tot10} - n^*_{10})}{N^*_{tot10}(n^*_{10} - 1)} , \quad \text{and} \quad FPC_{01} = \frac{n^*_{01}(N^*_{tot01} - n^*_{01})}{N^*_{tot01}(n^*_{01} - 1)} .$$

In these FPC definitions, the terms are again defined based on the <u>original</u> cell counts in Table 1:

$N^*_{tot11} = N^*_{tot10} = n_1 + n_2 + n_3 + n_4$, $N^*_{tot01} = n_5 + n_6 + n_7$, $n^*_{11} = n_1 + n_2$, $n^*_{10} = n_3 + n_4$, and

$n^*_{01} = n_5 + n_6$. The rationale behind the FPC adjustments is the notion that $\bar{x}_{11}$, $\bar{x}_{10}$, and $\bar{x}_{01}$ effectively result from random samples of finite populations of size $N^*_{tot11}$, $N^*_{tot10}$, and $N^*_{tot01}$.

## Appendix C: Numerical Example

For illustration and the reproducibility of calculations for case count estimation, we simulated a single set of data under the anchor stream design based on one of the settings ($N_{tot} = 500$, $\psi = 0.1$, prevalence ($p$) = 0.2, $N = 100$ cases) summarized in Table 3. The observed cell counts (see Table 1) are as follows: $n_1 = 6$, $n_2 = 5$, $n_3 = 100$, $n_4 = 46$, $n_5 = 33$, $n_6 = 6$, and $n_7 = 304$. The results are displayed in **Table C.1**.

This example follows the template for the anchor stream design, in the sense that the prescribed number selected randomly into Stream 2 ($n_1+n_2+n_5+n_6 = 50$) is relatively small compared to the number ($n_1+n_2+n_3+n_4 = 157$) voluntarily tested in Stream 1. Note that the estimated prevalence based only on Stream 1 [$(n_2+n_4)/157 = 0.325$] corresponds to a clear overestimate of $N$ ($162.4 = 0.325 \times 500$), driven by the preferential sampling of those with symptoms (who are at high disease risk). In contrast, the random sample in Stream 2 yields an estimated prevalence [$(n_2+n_6)/50 = 0.22$] and case count ($110 = 0.22 \times 500$) that are much closer to the truth. The anchor stream point estimates are similar, but their precision benefits are reflected in the smaller SEs and narrower intervals shown in Table C.1. With regard to the latter, note that the credible interval reported with the preferred estimate $\hat{N}_{\hat{\psi}*}$ [eqn. (4)] differs somewhat from that reported with $\hat{N}_\psi$ [eqn. (2)]. As the estimated prevalence is in excess of the cut-off of 0.2 ($\hat{p}_{\hat{\psi}*} = 103.8/500 = 0.208$), it follows that the adjusted credible interval is reported in conjunction with $\hat{N}_{\hat{\psi}*}$ (see Appendix A).

For a complete treatment of this example, including R code to facilitate the calculations for case count estimation as well as data, code and results pertaining to estimating the mean of a continuous biomarker level (X) based on the anchor stream design, we refer readers to the following GitHub site: https://github.com/YZHA-yuzi/Anchor-Stream-Design.

**Table C.1. Numerical Example: Case Count Estimates with $N_{tot} = 500$ and $N = 100$**

| Estimator | Mean (SE) | 95% Credible Interval for N |
|---|---|---|
| $\hat{N}_{RS}$ [a] | 110.0 (28.1) | **(63.5, 171.5)** |
| $\hat{N}_{\psi}$ [b] | 111.0 (23.2) | **(76.8, 167.9)** |
| $\hat{N}_{\hat{\psi}*}$ [c] | 103.8 (21.9) | **(72.3, 164.4)** |

a  SE based on (1); FPC-adjusted Jeffreys prior-based credible interval (**bold**; see Appendix A)
b  SE based on (3); Unadjusted Dirichlet-multinomial-based credible interval (**bold**; see Appendix A)
c  SE based on (5); Adjusted Dirichlet-multinomial-based credible interval (**bold**; see Appendix A)

# Online Appendix: Expanded Simulation Results

Table S1. Simulations Evaluating Case Count Estimates with $N_{tot} = 50$

| | True Disease Prevalence (p) = 0.1 | | | |
|---|---|---|---|---|
| | $\psi = 0.2$ | | $\psi = 0.4$ | |
| Estimator | Mean (SD) [avg. SE] | CI coverage [avg. width] | Mean (SD) [avg. SE] | CI coverage [avg. width] |
| $\hat{N}_{RS}$ [a] | 5.0 (4.3) [4.5] | 95.2% [14.2] | 5.0 (2.6) [2.6] | 92.2% [8.4] |
| $\hat{N}_{\psi}$ [b] | 5.0 (3.0) [3.8] | 100.0% [9.2] | 5.0 (1.8) [1.9] | 92.6% [5.2] |
| $\hat{N}_{\hat{\psi}*}$ [c] | 5.0 (3.0) [2.7] | 94.5% [10.4] | 5.0 (1.8) [1.6] | 91.7% [5.6] |
| | True Disease Prevalence (p) = 0.2 | | | |
| $\hat{N}_{RS}$ [a] | 10.1 (5.7) [5.7] | 98.2% [17.2] | 10.0 (3.5) [3.5] | 93.2% [10.7] |
| $\hat{N}_{\psi}$ [b] | 10.1 (4.1) [4.5] | 98.3% [12.5] | 10.0 (2.5) [2.5] | 90.2% [7.7] |
| $\hat{N}_{\hat{\psi}*}$ [c] | 10.1 (4.1) [3.7] | 95.0% [14.0] | 10.0 (2.5) [2.2] | 90.6% [8.2] |

a  SE based on (1); The proposed FPC-adjusted Jeffreys prior-based credible interval is evaluated (see Appendix A)
b  SE based on (3); The proposed unadjusted Dirichlet-multinomial-based credible interval is evaluated (see Appendix A)
c  SE based on (5); The proposed adjusted Dirichlet-multinomial-based credible interval is evaluated (see Appendix A)

Note: If a calculated lower bound is lower than the total number of cases identified in Streams 1 and 2, that lower bound is set to be the total number of identified cases.

Table S2. Simulations Evaluating Case Count Estimates with $N_{tot} = 100$

| Estimator | True Disease Prevalence (p) = 0.05 | | | |
| --- | --- | --- | --- | --- |
| | $\psi = 0.1$ | | $\psi = 0.2$ | |
| | Mean (SD) [avg. SE] | CI coverage [avg. width] | Mean (SD) [avg. SE] | CI coverage [avg. width] |
| $\hat{N}_{RS}$ [a] | 5.0 (6.5) [8.2] | 99.5% [26.5] | 5.0 (4.4) [4.6] | 94.8% [15.8] |
| $\hat{N}_{\psi}$ [b] | 5.0 (4.4) [7.3] | 100.0% [16.4] | 5.0 (3.0) [3.8] | 100.0% [9.2] |
| $\hat{N}_{\hat{\psi}*}$ [c] | 5.0 (4.5) [4.1] | 98.8% [19.4] | 5.0 (3.0) [2.7] | 99.4% [10.8] |
| | True Disease Prevalence (p) = 0.1 | | | |
| $\hat{N}_{RS}$ [a] | 9.9 (9.1) [9.6] | 99.2% [30.3] | 10.1 (6.0) [5.9] | 97.4% [19.7] |
| $\hat{N}_{\psi}$ [b] | 10.0 (6.2) [8.0] | 100.0% [19.8] | 10.1 (4.1) [4.5] | 98.3% [12.5] |
| $\hat{N}_{\hat{\psi}*}$ [c] | 10.0 (6.4) [5.8] | 95.6% [25.4] | 10.1 (4.2) [3.8] | 98.6% [15.4] |

a SE based on (1); The proposed FPC-adjusted Jeffreys prior-based credible interval is evaluated (see Appendix A)
b SE based on (3); The proposed unadjusted Dirichlet-multinomial-based credible interval is evaluated (see Appendix A)
c SE based on (5); The proposed adjusted Dirichlet-multinomial-based credible interval is evaluated (see Appendix A)
Note: If a calculated lower bound is lower than the total number of cases identified in Streams 1 and 2, that lower bound is set to be the total number of identified cases.

**Table S3. Simulations Evaluating Case Count Estimates with Small Prevalence and $N_{tot}$ = 250**

| Estimator | \multicolumn{2}{c}{$\psi = 0.1$} | \multicolumn{2}{c}{$\psi = 0.2$} | \multicolumn{2}{c}{$\psi = 0.5$} |
|---|---|---|---|---|---|---|
| | Mean (SD) [avg. SE] | CI coverage [avg. width] | Mean (SD) [avg. SE] | CI coverage [avg. width] | Mean (SD) [avg. SE] | CI coverage [avg. width] |
| \multicolumn{7}{c}{True Disease Prevalence (p) = 0.05} |
| $\hat{N}_{RS}$ [a] | 12.0 (10.1) [10.3] | 97.8% [36.2] | 11.9 (6.8) [6.6] | 91.5% [22.9] | 12.0 (3.4) [3.4] | 96.6% [11.0] |
| $\hat{N}_{\psi}$ [b] | 12.0 (6.9) [8.3] | 99.2% [28.3] | 11.9 (4.6) [4.7] | 98.2% [16.8] | 12.0 (2.3) [2.2] | 92.2% [7.7] |
| $\hat{N}_{\hat{\psi}*}$ [c] | 12.0 (6.9) [6.3] | 99.2% [28.3] | 11.9 (4.6) [4.1] | 98.2% [16.8] | 12.0 (2.3) [2.1] | 92.2% [7.7] |
| \multicolumn{7}{c}{True Disease Prevalence (p) = 0.1} |
| $\hat{N}_{RS}$ [a] | 25.0 (14.3) [13.9] | 90.9% [46.8] | 24.9 (9.5) [9.3] | 93.9% [31.5] | 24.9 (4.7) [4.7] | 96.9% [16.0] |
| $\hat{N}_{\psi}$ [b] | 25.2 (9.9) [10.3] | 98.4% [39.6] | 25.0 (6.6) [6.5] | 94.9% [25.3] | 25.0 (3.3) [3.3] | 95.6% [12.5] |
| $\hat{N}_{\hat{\psi}*}$ [c] | 25.1 (9.9) [9.0] | 98.0% [39.4] | 25.0 (6.5) [6.0] | 94.9% [25.3] | 25.0 (3.2) [3.1] | 95.6% [12.5] |

a SE based on (1); The proposed FPC-adjusted Jeffreys prior-based credible interval is evaluated (see Appendix A)
b SE based on (3); The proposed unadjusted Dirichlet-multinomial-based credible interval is evaluated (see Appendix A)
c SE based on (5); The proposed adjusted Dirichlet-multinomial-based credible interval is evaluated (see Appendix A)

Note: If a calculated lower bound is lower than the total number of cases identified in Streams 1 and 2, that lower bound is set to be the total number of identified cases.

**Table S4. Simulations Evaluating Case Count Estimates with Higher Prevalence and $N_{tot}$ = 250**

| Estimator | \multicolumn{2}{c}{$\psi = 0.1$} | | \multicolumn{2}{c}{$\psi = 0.2$} | | \multicolumn{2}{c}{$\psi = 0.5$} |
|---|---|---|---|---|---|---|---|
| | Mean (SD) [avg. SE] | CI coverage [avg. width] | Mean (SD) [avg. SE] | CI coverage [avg. width] | Mean (SD) [avg. SE] | CI coverage [avg. width] |
| \multicolumn{7}{c}{True Disease Prevalence (p) = 0.2} |
| $\hat{N}_{RS}$ [a] | 49.8 (19.0) [18.7] | 94.2% [61.6] | 49.8 (12.6) [12.6] | 95.0% [43.4] | 49.8 (6.3) [6.3] | 96.1% [22.7] |
| $\hat{N}_{\psi}$ [b] | 49.9 (13.5) [13.7] | 95.3% [55.1] | 49.8 (9.0) [9.2] | 96.0% [36.3] | 49.9 (4.5) [4.7] | 95.7% [18.4] |
| $\hat{N}_{\hat{\psi}*}$ [c] | 49.9 (13.4) [12.5] | 94.3% [53.4] | 49.8 (8.8) [8.6] | 95.1% [35.7] | 49.9 (4.4) [4.4] | 95.2% [18.0] |
| \multicolumn{7}{c}{True Disease Prevalence (p) = 0.5} |
| $\hat{N}_{RS}$ [a] | 124.8 (23.8) [23.8] | 96.6% [83.3] | 125.0 (16.0) [15.8] | 95.8% [59.4] | 125.0 (8.0) [7.9] | 95.8% [30.6] |
| $\hat{N}_{\psi}$ [b] | 124.7 (19.5) [22.0] | 97.4% [87.5] | 124.9 (13.2) [14.9] | 97.3% [58.3] | 125.0 (6.6) [7.5] | 97.3% [29.3] |
| $\hat{N}_{\hat{\psi}*}$ [c] | 124.9 (17.9) [18.0] | 94.7% [77.2] | 125.0 (11.9) [12.2] | 95.3% [51.2] | 125.0 (6.0) [6.2] | 95.7% [25.4] |

a SE based on (1); The proposed FPC-adjusted Jeffreys prior-based credible interval is evaluated (see Appendix A)

b SE based on (3); The proposed unadjusted Dirichlet-multinomial-based credible interval is evaluated (see Appendix A)

c SE based on (5); The proposed adjusted Dirichlet-multinomial-based credible interval is evaluated (see Appendix A)

Note: If a calculated lower bound is lower than the total number of cases identified in Streams 1 and 2, that lower bound is set to be the total number of identified cases.

**Table S5. Simulations Evaluating Case Count Estimates with Small Prevalence and $N_{tot}$ = 1000**

| | \multicolumn{2}{c|}{$\psi = 0.1$} | \multicolumn{2}{c|}{$\psi = 0.2$} | \multicolumn{2}{c|}{$\psi = 0.5$} |
|---|---|---|---|---|---|---|
| Estimator | Mean (SD) [avg. SE] | CI coverage [avg. width] | Mean (SD) [avg. SE] | CI coverage [avg. width] | Mean (SD) [avg. SE] | CI coverage [avg. width] |
| \multicolumn{7}{c}{True Disease Prevalence (p) = 0.05} |
| $\hat{N}_{RS}$ [a] | 49.7 (20.6) [20.2] | 94.8% [70.4] | 49.8 (13.8) [13.6] | 95.6% [48.2] | 49.9 (7.0) [6.9] | 94.1% [24.7] |
| $\hat{N}_{\psi}$ [b] | 49.7 (14.0) [13.6] | 94.5% [54.7] | 49.9 (9.2) [9.2] | 95.4% [36.3] | 49.9 (4.7) [4.7] | 94.9% [18.4] |
| $\hat{N}_{\hat{\psi}*}$ [c] | 49.7 (14.0) [12.9] | 94.5% [54.7] | 49.9 (9.2) [8.9] | 95.4% [36.3] | 49.9 (4.7) [4.6] | 94.9% [18.4] |
| \multicolumn{7}{c}{True Disease Prevalence (p) = 0.1} |
| $\hat{N}_{RS}$ [a] | 100.0 (28.5) [28.2] | 94.9% [100.4] | 100.1 (19.2) [18.9] | 95.1% [70.2] | 100.1 (9.4) [9.5] | 95.5% [36.2] |
| $\hat{N}_{\psi}$ [b] | 100.1 (19.8) [19.6] | 95.1% [78.2] | 100.1 (13.3) [13.3] | 94.9% [52.0] | 100.0 (6.6) [6.7] | 95.7% [26.2] |
| $\hat{N}_{\hat{\psi}*}$ [c] | 100.1 (19.6) [18.7] | 95.1% [78.2] | 100.1 (13.2) [12.8] | 94.9% [52.0] | 100.0 (6.5) [6.5] | 95.7% [26.2] |

a SE based on (1); The proposed FPC-adjusted Jeffreys prior-based credible interval is evaluated (see Appendix A)

b SE based on (3); The proposed unadjusted Dirichlet-multinomial-based credible interval is evaluated (see Appendix A)

c SE based on (5); The proposed adjusted Dirichlet-multinomial-based credible interval is evaluated (see Appendix A)

Note: If a calculated lower bound is lower than the total number of cases identified in Streams 1 and 2, that lower bound is set to be the total number of identified cases.

**Table S6. Simulations Evaluating Case Count Estimates with Higher Prevalence and $N_{tot}$ = 1000**

| | True Disease Prevalence (p) = 0.2 | | | | | |
|---|---|---|---|---|---|---|
| | $\psi = 0.1$ | | $\psi = 0.2$ | | $\psi = 0.5$ | |
| Estimator | Mean (SD) [avg. SE] | CI coverage [avg. width] | Mean (SD) [avg. SE] | CI coverage [avg. width] | Mean (SD) [avg. SE] | CI coverage [avg. width] |
| $\hat{N}_{RS}$ [a] | 199.9 (37.7) [37.9] | 95.3% [141.4] | 199.9 (25.2) [25.3] | 95.3% [97.7] | 199.8 (12.7) [12.6] | 95.0% [49.4] |
| $\hat{N}_{\psi}$ [b] | 200.0 (27.3) [28.1] | 95.8% [111.0] | 200.0 (18.2) [18.9] | 96.0% [74.0] | 200.0 (9.0) [9.5] | 96.2% [37.1] |
| $\hat{N}_{\hat{\psi}*}$ [c] | 199.9 (26.6) [26.4] | 95.8% [110.0] | 200.0 (17.7) [17.8] | 95.9% [72.9] | 200.0 (8.8) [9.0] | 95.9% [36.4] |
| | True Disease Prevalence (p) = 0.5 | | | | | |
| $\hat{N}_{RS}$ [a] | 499.6 (47.4) [47.5] | 95.3% [183.7] | 499.5 (31.6) [31.6] | 95.1% [123.3] | 499.8 (15.8) [15.8] | 95.0% [61.9] |
| $\hat{N}_{\psi}$ [b] | 499.7 (39.4) [44.8] | 97.5% [176.0] | 499.7 (26.5) [29.9] | 97.6% [117.2] | 499.9 (13.2) [15.0] | 97.5% [58.7] |
| $\hat{N}_{\hat{\psi}*}$ [c] | 499.5 (35.3) [37.0] | 96.1% [154.3] | 499.7 (23.8) [24.8] | 96.2% [102.3] | 499.9 (11.9) [12.4] | 96.1% [51.1] |

a  SE based on (1); The proposed FPC-adjusted Jeffreys prior-based credible interval is evaluated (see Appendix A)
b  SE based on (3); The proposed unadjusted Dirichlet-multinomial-based credible interval is evaluated (see Appendix A)
c  SE based on (5); The proposed adjusted Dirichlet-multinomial-based credible interval is evaluated (see Appendix A)

Note: If a calculated lower bound is lower than the total number of cases identified in Streams 1 and 2, that lower bound is set to be the total number of identified cases.

**Table S7. Simulations Evaluating Mean Estimates for Continuous X with $N_{tot}$ = 250 and $\psi$ = 0.2**

| | Overall Mean Estimators | | | | | |
|---|---|---|---|---|---|---|
| | **p = 0.1, True $\mu_X$ =1.785** | | **p = 0.2, True $\mu_X$ =2.42** | | **p = 0.5, True $\mu_X$ =4.325** | |
| Estimator | Mean (SD) [avg. SE] | CI coverage [avg. width] | Mean (SD) [avg. SE] | CI coverage [avg. width] | Mean (SD) [avg. SE] | CI coverage [avg. width] |
| $\bar{x}_{1\bullet}$ [a] | 2.906 (0.31) [--] | -- | 4.068 (0.36) [--] | -- | 6.596 (0.34) [--] | -- |
| $\bar{x}_{2\bullet}$ [b] | 1.783 (0.34) [0.32] | 92.4% [1.25] | 2.423 (0.41) [0.40] | 93.4% [1.55] | 4.333 (0.51) [0.49] | 93.4% [1.91] |
| $\hat{\mu}_X$ [c] | 1.786 (0.23) [0.21] | 92.3% [0.83] | 2.419 (0.25) [0.25] | 93.1% [0.96] | 4.330 (0.30) [0.29] | 94.0% [1.13] |
| | Mean Estimators for Cases | | | | | |
| | **p = 0.1, True $\mu_{X,cases}$ =7.5** | | **p = 0.2, True $\mu_{X,cases}$ =7.5** | | **p = 0.5, True $\mu_{X,cases}$ =7.5** | |
| $\bar{x}_{1\bullet,cases}$ [d] | 9.088 (0.57) [0.54] | 29.2% [2.11] | 9.086 (0.40) [0.39] | 6.1% [1.52] | 9.089 (0.25) [0.25] | 0.0% [0.96] |
| $\bar{x}_{2\bullet,cases}$ [e] | 7.486 (1.27) [1.08] | 84.6% [4.12] | 7.507 (0.85) [0.83] | 93.0% [3.26] | 7.501 (0.52) [0.52] | 94.4% [2.02] |
| $\hat{\mu}_{X,cases}$ [f] | 7.489 (0.76) [0.66] | 93.5% [2.52] | 7.555 (0.60) [0.54] | 94.0% [2.11] | 7.515 (0.36) [0.35] | 93.5% [1.35] |
| | Mean Estimators for Non-Cases | | | | | |
| | **p=0.1, True $\mu_{X,noncases}$ =1.15** | | **p=0.2, True $\mu_{X,noncases}$ =1.15** | | **p=0.5, True $\mu_{X,noncases}$ =1.15** | |
| $\bar{x}_{1\bullet,noncases}$ | 1.504 (0.20) [0.20] | 57.8% [0.79] | 1.500 (0.22) [0.21] | 61.5% [0.83] | 1.501 (0.27) [0.27] | 73.2% [1.06] |
| $\bar{x}_{2\bullet,noncases}$ | 1.151 (0.23) [0.23] | 94.2% [0.89] | 1.149 (0.24) [0.24] | 93.9% [0.94] | 1.154 (0.31) [0.31] | 93.5% [1.20] |
| $\hat{\mu}_{X,noncases}$ | 1.151 (0.20) [0.20] | 94.0% [0.77] | 1.150 (0.21) [0.21] | 94.2% [0.82] | 1.156 (0.27) [0.26] | 93.4% [1.03] |
| | Estimators for Case vs. Non-Case Mean Difference | | | | | |
| | **p=0.1, True $\mu_{diff}$ =6.35** | | **p=0.2, True $\mu_{diff}$ =6.35** | | **p=0.5, True $\mu_{diff}$ =6.35** | |
| $\bar{x}_{1\bullet,diff}$ | 7.584 (0.60) [0.58] | 49.2% [2.27] | 7.587 (0.45) [0.45] | 26.4% [1.74] | 7.588 (0.37) [0.37] | 9.5% [1.44] |
| $\bar{x}_{2\bullet,diff}$ | 6.335 (1.30) [1.12] | 84.7% [4.28] | 6.358 (0.89) [0.87] | 93.3% [3.40] | 6.347 (0.60) [0.60] | 94.6% [2.35] |
| $\hat{\mu}_{X,diff}$ | 6.337 (0.79) [0.70] | 93.3% [2.66] | 6.404 (0.64) [0.59] | 94.2% [2.28] | 6.359 (0.46) [0.44] | 94.1% [1.73] |

a Estimated mean for individuals sampled in Stream 1; SE and CIs not reported

b Estimated mean for individuals sampled in Stream 2; SE incorporates FPC adjustment with Wald-type CIs; SE of $\bar{x}_{2\bullet}$ equals sample standard deviation of X among those sampled in Stream 2 divided by square root of the number of individuals sampled in Stream 2

c SE based on bootstrap with percentile CIs incorporating FPC adjustments (see Appendix B)

d Estimated mean among cases sampled in Stream 1; SE based on bootstrap with percentile CIs

e Estimated mean among cases sampled in Stream 2; SE based on bootstrap with percentile CIs

f SE based on bootstrap with percentile CIs (see Appendix B)

**Table S8. Simulations Evaluating Mean Estimates for Continuous X with $N_{tot}$ = 1000 and $\psi$ =0.2**

| | Overall Mean Estimators | | | | | |
|---|---|---|---|---|---|---|
| | **p = 0.1, True $\mu_X$ =1.785** | | **p = 0.2, True $\mu_X$ =2.42** | | **p = 0.5, True $\mu_X$ =4.325** | |
| Estimator | Mean (SD) [avg. SE] | CI coverage [avg. width] | Mean (SD) [avg. SE] | CI coverage [avg. width] | Mean (SD) [avg. SE] | CI coverage [avg. width] |
| $\bar{x}_{1\bullet}$ [a] | 2.901 (0.15) [--] | -- | 4.063 (0.18) [--] | -- | 6.592 (0.17) [--] | -- |
| $\bar{x}_{2\bullet}$ [b] | 1.785 (0.17) [0.16] | 93.2% [0.63] | 2.423 (0.20) [0.20] | 93.9% [0.77] | 4.327 (0.25) [0.24] | 94.0% [0.95] |
| $\hat{\mu}_X$ [c] | 1.786 (0.11) [0.11] | 93.5% [0.41] | 2.420 (0.13) [0.12] | 93.6% [0.48] | 4.325 (0.15) [0.14] | 94.4% [0.56] |
| | Mean Estimators for Cases | | | | | |
| | **p = 0.1, True $\mu_{X,cases}$ =7.5** | | **p = 0.2, True $\mu_{X,cases}$ =7.5** | | **p = 0.5, True $\mu_{X,cases}$ =7.5** | |
| $\bar{x}_{1\bullet,cases}$ [d] | 9.090 (0.28) [0.28] | 0.1% [1.08] | 9.091 (0.20) [0.20] | 0.0% [0.76] | 9.092 (0.12) [0.12] | 0.0% [0.48] |
| $\bar{x}_{2\bullet,cases}$ [e] | 7.502 (0.59) [0.58] | 94.2% [2.28] | 7.508 (0.41) [0.41] | 94.3% [1.60] | 7.500 (0.26) [0.26] | 95.0% [1.01] |
| $\hat{\mu}_{X,cases}$ [f] | 7.537 (0.42) [0.42] | 93.9% [1.62] | 7.516 (0.30) [0.29] | 93.9% [1.14] | 7.503 (0.18) [0.17] | 94.0% [0.67] |
| | Mean Estimators for Non-Cases | | | | | |
| | **p=0.1, True $\mu_{X,noncases}$ =1.15** | | **p=0.2, True $\mu_{X,noncases}$ =1.15** | | **p=0.5, True $\mu_{X,noncases}$ =1.15** | |
| $\bar{x}_{1\bullet,noncases}$ | 1.500 (0.10) [0.10] | 7.0% [0.39] | 1.501 (0.11) [0.11] | 9.4% [0.42] | 1.498 (0.14) [0.14] | 27.5% [0.53] |
| $\bar{x}_{2\bullet,noncases}$ | 1.151 (0.11) [0.11] | 94.9% [0.45] | 1.150 (0.12) [0.12] | 94.4% [0.47] | 1.147 (0.15) [0.15] | 94.6% [0.60] |
| $\hat{\mu}_{X,noncases}$ | 1.151 (0.10) [0.10] | 94.6% [0.39] | 1.150 (0.11) [0.11] | 94.1% [0.41] | 1.148 (0.13) [0.13] | 94.2% [0.52] |
| | Estimators for Case vs. Non-Case Mean Difference | | | | | |
| | **p=0.1, True $\mu_{diff}$ =6.35** | | **p=0.2, True $\mu_{diff}$ =6.35** | | **p=0.5, True $\mu_{diff}$ =6.35** | |
| $\bar{x}_{1\bullet,diff}$ | 7.590 (0.30) [0.29] | 3.1% [1.15] | 7.590 (0.22) [0.22] | 0.1% [0.87] | 7.594 (0.19) [0.18] | 0.0% [0.72] |
| $\bar{x}_{2\bullet,diff}$ | 6.351 (0.60) [0.59] | 94.2% [2.33] | 6.358 (0.43) [0.43] | 94.4% [1.67] | 6.353 (0.30) [0.30] | 94.8% [1.17] |
| $\hat{\mu}_{X,diff}$ | 6.386 (0.44) [0.43] | 94.2% [1.68] | 6.366 (0.32) [0.31] | 93.9% [1.22] | 6.355 (0.23) [0.22] | 94.3% [0.87] |

a Estimated mean for individuals sampled in Stream 1; SE and CIs not reported

b Estimated mean for individuals sampled in Stream 2; SE incorporates FPC adjustment with Wald-type CIs; SE of $\bar{x}_{2\bullet}$ equals sample standard deviation of X among those sampled in Stream 2 divided by square root of the number of individuals sampled in Stream 2

c SE based on bootstrap with percentile CIs incorporating FPC adjustments (see Appendix B)

d Estimated mean among cases sampled in Stream 1; SE based on bootstrap with percentile CIs

e Estimated mean among cases sampled in Stream 2; SE based on bootstrap with percentile CIs

f SE based on bootstrap with percentile CIs (see Appendix B)